# An Extremely Elongated Cloud over Arsia Mons Volcano on Mars: I. Life Cycle

J. Hernández-Bernal[1,2], A. Sánchez-Lavega[1], T. del Río-Gaztelurrutia[1], E. Ravanis[3], A. Cardesín-Moinelo[3,4], K. Connour[5], D. Tirsch[6], I. Ordóñez-Etxeberria[1], B. Gondet[7], S. Wood[8], D. Titov[9], N. M. Schneider[5], R. Hueso[1], R. Jaumann[10], E. Hauber[6]

[1]Dpto. Física Aplicada I, EIB, Universidad País Vasco UPV/EHU, Bilbao, Spain [2]Aula EspaZio Gela, Escuela de Ingeniería de Bilbao, Universidad del País Vasco UPV/EHU, Bilbao, Spain [3]European Space Agency, ESAC, Madrid, Spain [4]Instituto de Astrofísica e Ciências do Espaço, Obs. Astronomico de Lisboa, Portugal [5]Laboratory for Atmospheric and Space Physics, University of Colorado, Boulder, USA [6]German Aerospace Center (DLR), Institute of Planetary Research, Berlin, Germany. [7]Institut d'Astrophysique Spatiale, CNRS/University Paris Sud, Orsay, France [8]European Space Agency, ESOC, Darmstadt, Germany [9]European Space Agency, ESTEC, Noordwijk, The Netherlands [10]Freie Universitaet Berlin, Institute of Geological Sciences, Berlin, Germany

**Corresponding author:** Jorge Hernández-Bernal (jorge.hernandez@ehu.eus)

**Key Points:**

- We report a new phenomenon consisting of an extremely elongated water ice cloud (up to 1800 km) extending westward from the Arsia Mons volcano.
- The cloud reaches the mesosphere (45 km), and expands at a velocity of around 170 m/s in Martian Year 34.
- This cloud repeatedly forms in the early mornings, and repeats in a daily cycle between Ls 220º and 320º every martian year.

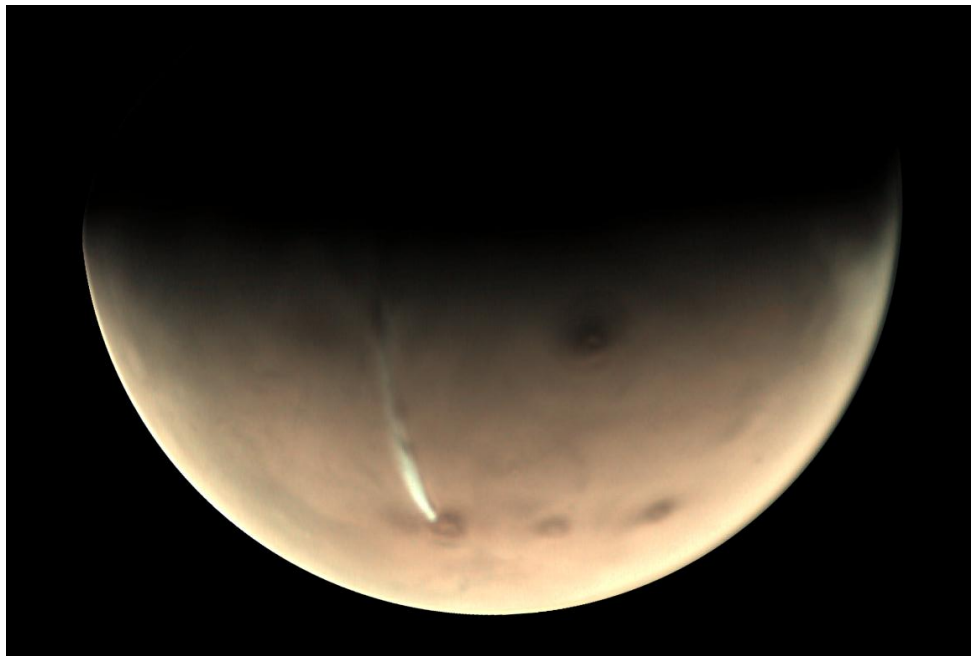

**Figure 1**. The elongated cloud extending around 1800 km to the west from Arsia Mons volcano on October 10 2018 (MY34). Image acquired by the VMC camera on board MEX. ESA press release:
http://www.esa.int/Science_Exploration/Space_Science/Mars_Express/Mars_Express_keeps_an_eye_on_curious_cloud

**Abstract**

We report a previously unnoticed annually repeating phenomenon consisting of the daily formation of an extremely elongated cloud extending as far as 1800 km westward from Arsia Mons. It takes place in the Solar Longitude (Ls) range of ~220º-320º, around the Southern solstice. We study this Arsia Mons Elongated Cloud (AMEC) using images from different orbiters, including ESA Mars Express, NASA MAVEN, Viking 2, MRO, and ISRO Mars Orbiter Mission (MOM). We study the AMEC in detail in Martian Year (MY) 34 in terms of Local Time and Ls and find that it exhibits a very rapid daily cycle: the cloud growth starts before sunrise on the western slope of the volcano, followed by a westward expansion that lasts 2.5 hours with a velocity of around 170 m/s in the mesosphere (~45 km over the areoid). The cloud formation then ceases, it detaches from its formation point, and continues moving westward until it evaporates before the afternoon, when most sun-synchronous orbiters observe. Moreover we comparatively study observations from different years (i.e. MYs 29-34) in search of interannual variations and find that in MY33 the cloud exhibits lower activity, whilst in MY34 the beginning of its formation was delayed compared to other years, most likely due to the Global Dust Storm. This phenomenon takes place in a season known for the general lack of clouds on Mars. In this paper we focus on observations, and a theoretical interpretation will be the subject of a separate paper.

**Plain Language Summary**

In September and October 2018 the Visual Monitoring Camera on board Mars Express observed a spectacular water ice cloud extending as far as 1800 km westward from the Arsia Mons Volcano on Mars. This curious extremely elongated cloud caught the attention of the public (http://www.esa.int/Science_Exploration/Space_Science/Mars_Express/Mars_Express_keeps_an_eye_on_curious_cloud). We study this Arsia Mons Elongated Cloud (AMEC) with the aid of several instruments orbiting Mars. We find that the AMEC repeated regularly each morning for a number of months, and that it is an annually-repeating phenomenon that takes place every Martian Year around the southern hemisphere spring and summer. The AMEC follows a rapid daily cycle: it starts to expand from Arsia Mons at dawn at an altitude of about ~45 km, and for ~2.5 hours it expands westward as fast as 170 m/s (around 600 km/h). The cloud then detaches from Arsia Mons and evaporates before noon. In previous Martian Years, few observations of this phenomenon are available because most cameras orbiting Mars are placed in orbits where they can only observe during the afternoon, whereas this cloud takes place in the early morning, when observational coverage is much lower.

## 1. Introduction

Water ice clouds are present throughout the Martian Year (MY) with large temporal and spatial variability (Wang & Ingersoll, 2002; Clancy et al., 2017; Wolff et al. 2019). They have been observed from ground-based telescopes (Herschel, 1784; Slipher, 1927; Smith & Smith 1972), orbiting spacecraft (Leovy et al., 1971;1973; Curran et al., 1973) and rovers (Kloos, 2016; Wolff et al., 2005).

Prominent clouds are commonly found in the region of Tharsis, and have been noticed since the telescopic era (Slipher, 1927; Smith & Smith, 1972). Based on early topographic maps, Sagan et al. (1971) proposed a topographic origin for these clouds over Tharsis and other high altitude regions. The observations of the Mariner and Viking missions supported this idea and confirmed the water ice composition (Peale et al., 1973; Curran et al., 1973). More recently Michaels et al. (2006) have shown in a mesoscale model how the circulation induced by giant volcanoes injects large amounts of aerosols into the global circulation.

Water ice clouds on Mars, including the topographic ones, are most commonly seen as part of the Aphelion Cloud Belt (ACB) in the season around Solar Longitude (Ls) ~40º-140º (Clancy et al., 1996; 2017; James et al., 1996). Benson et al. (2003, 2006) studied the seasonal distribution of clouds around the giant volcanoes on Mars using afternoon observations, and found that the volcanoes usually lack clouds during the dusty season, in the Ls range ~200º-360º. Arsia Mons is an exception to this, as it shows cloud activity in the Ls range 245º-322º, while in general little cloud activity is found on Mars in this season (Benson et al. 2003, 2006; Wang & Ingersoll, 2002).

There is a bias in the local time of observations made by orbiting spacecraft that could affect our knowledge of the life cycle of Martian water ice cloud systems, which is that most observations in the past two decades of Mars exploration have been acquired by spacecraft in afternoon sun-synchronous orbits (whose observations are centered at local afternoon). This is the case for the Mars Global Surveyor (MGS), Mars Odyssey (later shifted to 7AM-7PM local times; Smith, 2019a), and Mars Reconnaissance Orbiter (MRO). Until 2014, when MAVEN (Mars Atmosphere and Volatile EvolutioN) and the Mars Orbiter Mission (MOM) arrived at Mars, Mars Express (MEX) was the only mission in a non-sun-synchronous orbit. The ExoMars Trace Gas Orbiter (TGO), which arrived in 2016 and began science operations in 2018, is also in a non-sun-synchronous orbit.

Given this observational bias, few authors have researched the daily evolution of water ice clouds. Hunt et al. (1980) studied the daily evolution of several orographic clouds based on Viking images, and Akabane et al. (2002) used ground based telescopic observations (see also Glenar et al., 2003). Other authors (e.g. Wilson et al., 2007) took advantage of the night side observations of afternoon sun-synchronous missions to analyze the behavior of clouds during the night. More recently, Smith et al. (2019a) compared THEMIS observations from before and after the shift in the local time of the Mars Odyssey sun-synchronous orbit. Afternoon sun-synchronous orbiters can only be used to study variations within a few hours around the afternoon, given the narrow range of different local times covered by their field of view (Wolff et al., 2019; Benson et al., 2003,2006). Landers and rovers are also suitable observers of the diurnal variations (Kloos et al., 2018; Sutton et al., 1978). More observations from non-sun-synchronous orbiters (e.g. Giuranna et al., 2019) are necessary to achieve a full comprehension of the diurnal variability of the martian climate.

We report the observation of a previously observed but unnoticed phenomenon that is the formation of an extremely Elongated Cloud extending to the west from the Arsia Mons volcano (Figure 1). We first noticed this phenomenon in 2018 (MY34), as a result of the better observational conditions favoured by MEX and MAVEN orbits. We then found this phenomenon in archived images from previous years, always around the same season, in the Ls range 220º-320º, which included the Southern solstice (Ls 270º) and the perihelion (Ls 251º). This Arsia Mons Elongated Cloud (AMEC) develops in the early morning and dissipates before afternoon; this is why it was seldom observed by the instruments on missions in afternoon sun-synchronous orbits, such as the Mars Color Imager (MARCI) on MRO or Mars Orbiter Camera (MOC) on MGS.

The occurrence of the AMEC in this particular season is likely related to a specific coincidence of atmospheric conditions such as winds, temperatures, pressure, and water vapor content in the region. Since the cloud seems to occurevery MY, we are interested in the interannual variability of the phenomenon, and its possible value as a proxy for the general state of the atmosphere in different MYs (Sánchez-Lavega et al., 2018b). We note that 2018 observations of the AMEC were after the Martian Global Dust Storm of 2018 (GDS 2018; Sánchez-Lavega et al., 2019, Guzewich et al., 2019), and Global Dust Storms are the most conspicuous Martian events that exhibit interannual variability. In fact, Benson et al. (2006), based on afternoon observations, reported lower cloud activity over Arsia Mons following the 2001 GDS, compared to the previous MY.

In this first paper, we analyze observations of the Arsia Mons Elongated Cloud (AMEC). We start our analysis by focusing on MY 34 (which corresponds to observations in 2018), as most available observations of the AMEC correspond to this year. Then, we compare the characteristics of the AMEC in MY 34 with all available observations from previous years, in search of interannual variations. A theoretical analysis of the phenomenon will be the subject of a separate paper.

In section 2, we describe our Observations and Methods. In section 3, we briefly describe the Tharsis region and the topographic profiles of the four giant volcanoes, including Arsia Mons. In section 4, we describe the daily cycle followed by the AMEC, from twilight to noon and afternoon, as inferred from observations in MY34. In section 5, we show several morphologic and dynamic features of the cloud and describe its seasonal evolution. In section 6, we measure the altitude of the cloud. In section 7, we compare MY34 with observations from previous years, with a special interest in interannual variations. Finally, in section 8 we summarize our observations and present our conclusions.

**Table 1. Spacecraft and instruments temporal coverage**

| Instrument | Spacecraft | Number of observations | Spatial resolution (km px$^{-1}$) | MY |
|---|---|---|---|---|
| VMC | MEX | 63 | 12 | 29,30,31,33,34 |
| HRSC | MEX | 3 | 1 | 33,34 |
| OMEGA | MEX | 1 | 5 | 34 |
| IUVS | MAVEN | 29 | 10 | 34 |
| MCC | MOM | 3 | 1 | 32 |
| VIS | Viking 2 | 2 | 1 | 12 |
| MARCI | MRO | * | 1 | * |

**Notes to Table 1**: For each instrument, we specify the number of observations, the typical spatial resolution of the images in our dataset, and the MY. Refer to tables S1-S6 in the supporting material for further details. In the case of VMC, the actual number of images is much larger than the number of observations, since each observation consists of a number of similar images with different exposure times. * Since MARCI is used only for support and performs daily systematic observations, we do not indicate the number of observations or the MYs covered.

## 2. Observations and Methods

In this study, we have taken advantage of a variety of imaging systems onboard different spacecraft (Table 1 and Figure 3):

- VMC (Visual Monitoring Camera) (Ormston et al., 2011; Sánchez-Lavega et al. 2018a), on board MEX, is a low resolution RGB color camera that usually observes from the apoapsis of the MEX orbit at 10500km (Details on the orbit and VMC operations can be found in Ravanis et al., 2020). Images are usually taken in a series covering several minutes, enabling the measurement of atmospheric motions from the tracking of clouds and dust features (Hernández-Bernal et al., 2019).

- HRSC (High Resolution Stereo Camera) (Jaumann et al., 2007) on board MEX is a high resolution pushbroom camera that takes images in 9 different channels at different altitudes along the highly elliptical orbit.

- OMEGA (Observatoire pour la Minéralogie, l'Eau, les Glaces et l'Activité) (Bibring et al. 2004) on board MEX, is an imaging spectrometer that takes images simultaneously in several spectral bands.

- IUVS (Imaging Ultraviolet Spectrograph) (McClintock et al., 2015, Connour et al. 2019) on board MAVEN, is a scanning-slit spectrograph that takes composite false color UV images of Mars. IUVS acquires data during nearly every MAVEN orbit with a cadence of ~4.5 hours, enabling it to image recurring phenomena.

- MCC (Mars Colour Camera) (Arya et al. 2015) on board ISRO (Indian Space Research Organization) MOM, is an RGB color camera that acquires images at different altitudes from a highly elliptical orbit, usually with high resolution.

- VIS (Visual Imaging System) (Wellman et al., 1976) on board Viking 2 orbiter was a single channel visible camera of high resolution. Viking orbiters were in highly elliptical non sun-synchronous orbits.

- MARCI (Mars Color Imager) (Bell et al., 2009) on board MRO is a pushbroom camera that observes Mars during the afternoon from an afternoon sun-synchronous orbit, achieving high-resolution coverage of the whole planet every Martian sol. As MARCI only observes during the afternoon, it has never imaged the AMEC itself, therefore we use MARCI for support in the understanding of the AMEC daily cycle (section 4).

This includes all current missions in non sun-synchronous orbits, except TGO. We have excluded TGO since its only camera, CASSIS (Colour and Stereo Surface Imaging System), has a narrow field of view and thus it is not likely to have imaged the AMEC.

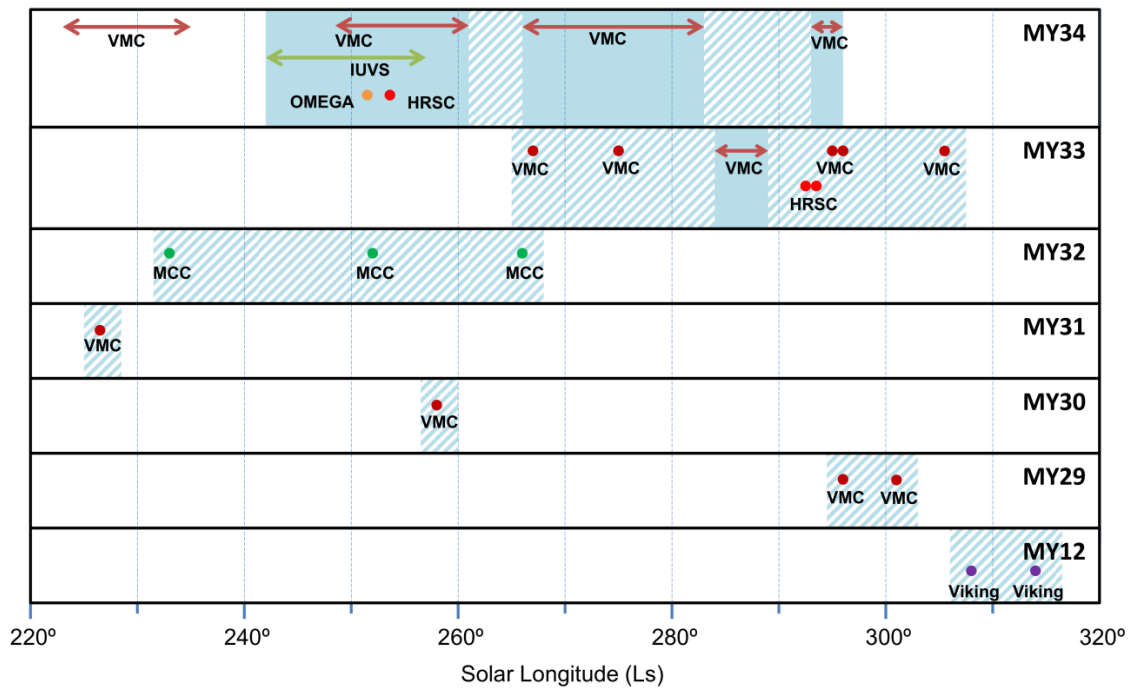

**Figure 2.** Seasonal coverage of Arsia Mons observations by different instruments. Colored points represent single observations and arrows represent periods with continuous observations (at least every few sols). Only observations between 5 and 10 LTST (Local True Solar Time) are included. Positive observations (i.e. with the AMEC present) are indicated by the background colour. Flat colored areas indicate continuous observations, and striped areas infer the probable minimum duration of the phenomenon. Single positive observations are indicated by a small striped area. The actual duration of the phenomenon is very likely longer than shown here for most MYs, as few suitable observations are available before MY34, and they are all positive. This figure will be discussed again in section 7 as part of the interannual comparison. Refer to tables S1-S6 in the supporting material for further details.

**An Extremely Elongated Cloud over Arsia Mons Volcano on Mars: I. Life Cycle.**
Hernández-Bernal et al. 2020. Manuscript accepted for publication on Journal of Geophysical Research
This document is distributed under CC BY-SA 3.0 IGO license

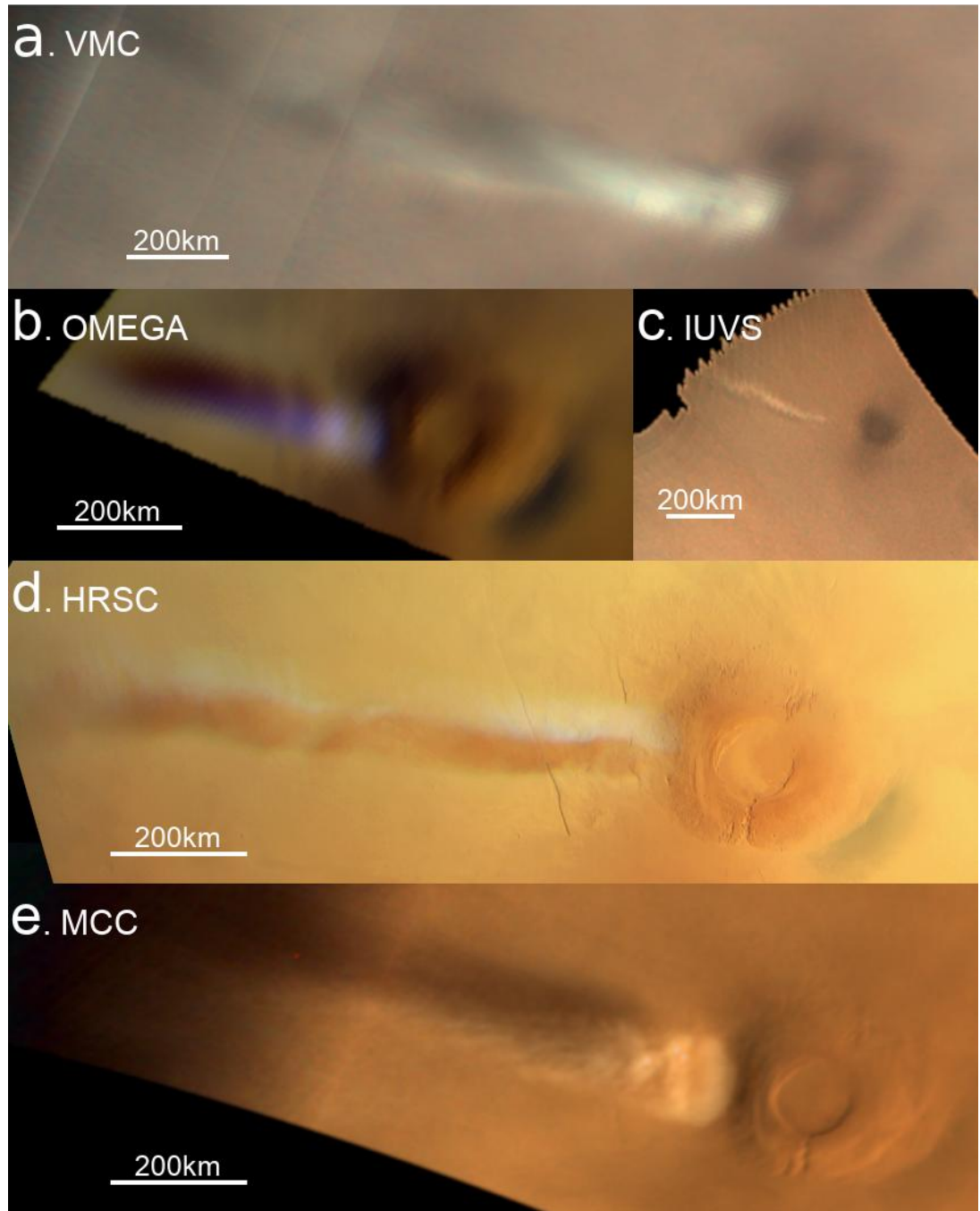

**Figure 3**. Examples of images from different intruments. Panels a to e show cylindrical projections of images displaying the AMEC at Arsia Mons. (a) VMC observation. 2018-10-22, MY 34, Ls 274º, LTST 7.8. (VMC observation 180371). (b) OMEGA observation. 2018-09-17, MY 34, Ls 252º, LTST 7.5. (MEX orbit 18612). (c) IUVS observation from apocenter. 2018-09-07, MY 34, Ls 245º, LTST 8.9. (MAVEN orbit 7673). (d) HRSC observation. 2018-09-21, MY 34, Ls 254º, LTST 8.5. (MEX orbit 18627). (e) MCC observation. 2014-12-13, MY 32, Ls 252º, LTST 6.7. (MCC product MCC_MRD_20141213T134802312_D_GDS)**.**

The largest set of observations corresponds to MY 34, spanning the orbital longitude range from Ls 223º to Ls 296º. This set comprises of about 75 observations obtained with the instruments VMC, IUVS, HRSC, and OMEGA, together with some MARCI observations. The use of this set of instruments has allowed us to follow the evolution of the phenomenon in MY34, both throughout the season and at a broad range of local times. There are far fewer observations from other MYs. Examples of images obtained by the various instruments used here are shown in Figure 3. Details on how the images from different instruments were processed and measured can be found in the supporting information. In general, we consider that a detection is positive if we find a morning cloud near the volcano or an elongated cloud west to the volcano, even if the volcano is not present in the image.

We used the geometry package Elkano (Hernández-Bernal et al., 2019, see for example Fig. 1) for the navigation of VMC, Viking/VIS, and MCC images, and for geometrical calculations in general. When necessary, spacecraft geometric information was extracted from SPICE (Acton 1996; Acton et al., 2018), with SPICE kernels provided by NASA NAIF in the case of MAVEN, MRO and Viking, and ESA SPICE Service in the case of MEX (Costa, 2018). In the case of MCC, we used kernels distributed by ISRO, and gaps in these kernels were covered through geometric information provided in the labels of MCC publicly available products. MARCI images were navigated and projected with the aid of the ISIS3 software package (Edwards, 1987). We projected the images onto cylindrical maps after correction of their luminosity (for uniformity near the terminator) before making measurements (see details in supporting material).

For topographic calculations (precise geometry calculations, and shadow measurements) we used Mars Orbiter Laser Altimeter (MOLA) data (Smith et al., 2001), and for topographic maps we use the topography produced by Fergason et al. (2018) from MOLA and HRSC data. All altitudes in this paper are given relative to the areoid as defined in MOLA data. For time calculations, we used algorithms and definitions described by Allison (1997) and Allison & McEwen (2000), and the enumeration of MYs proposed by Clancy et al. (2000). All Local True Solar Times (LTST) mentioned in this paper refer to the position of the volcanic rim of Arsia Mons volcano (239.5º E, 9.2º S), unless otherwise indicated.

Linear regressions shown in figures 6, 7 and 12 are calculated using orthogonal distance regression, which takes into account error in both variables.

## 3. Topography of the Tharsis region

The most prominent volcanic provinces on Mars are the Elysium region, Alba Mons, and the Tharsis Region, which contains the four tallest volcanoes on Mars: Olympus Mons, Ascraeus Mons, Pavonis Mons, and Arsia Mons (Figure 4). The entire Tharsis areais considerably higher than the mean altitude of the Martian surface.

Figure 4 shows a topographic map and the vertical profiles along the central axis of the four Tharsis volcanos. The summits of the four volcanos are ~8-20 km above their surrounding area, and at their base they extend ~300-500 km in diameter. All of the volcanoes also have depressed calderas in the range of ~50-100 km in diameter and 1-4 km in depth.

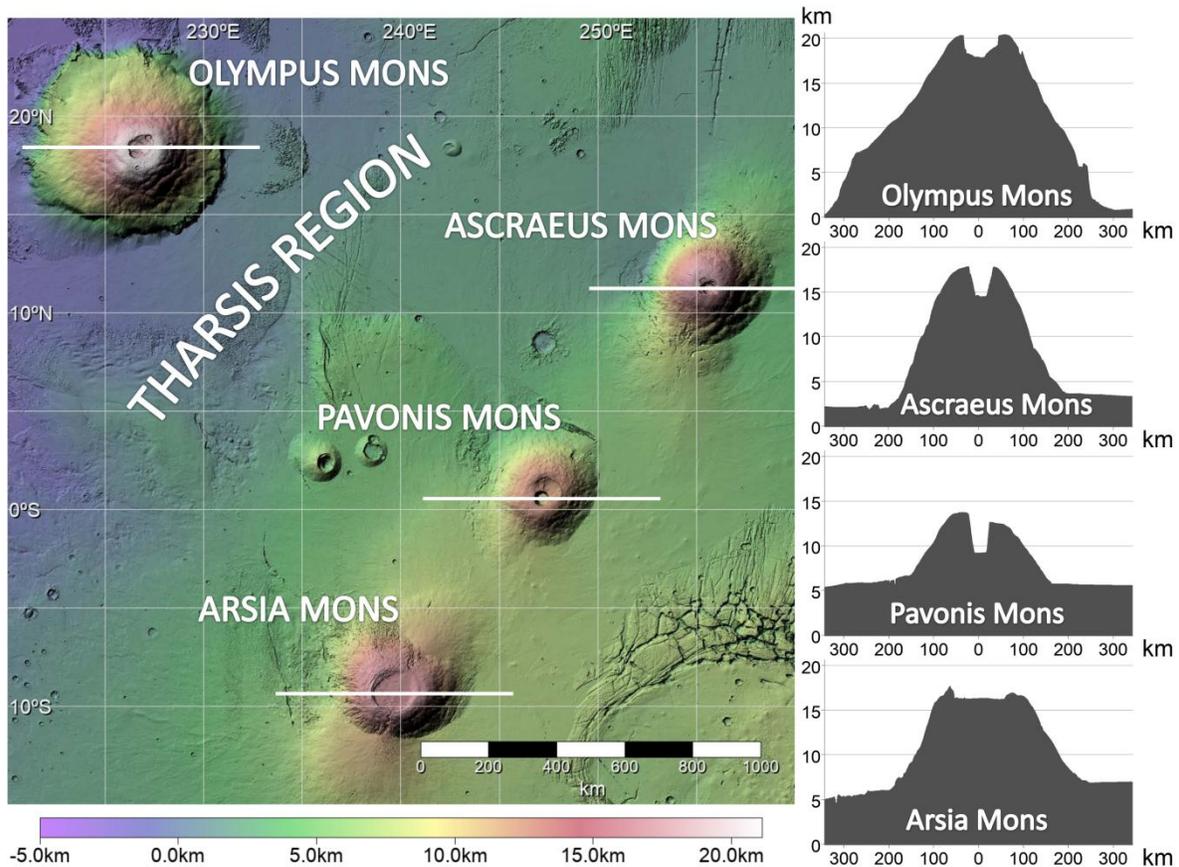

**Figure 4**. Topography of the Tharsis region from MOLA-HSRC data plot with hillshading (Fergason et al., 2018). The four giant volcanoes are apparent. Right panels show topographic profiles for each volcano from MOLA. Altitudes are relative to the areoid.

## 4 Daily cycle of the cloud in MY 34

The AMEC daily cycle was observed in detail in MY34. In the Ls range 242º-296º, the AMEC was always present in the early morning. The first observation of the AMEC was performed by IUVS on September 2, 2018 (Ls 242º). Prior to that date, the cloud was not present in suitable observations by VMC in August, in the Ls range 223-235º. The last image of the AMEC was obtained by VMC on November 28 (Ls 296º) and the region was not observed by VMC later in the season. Orbital and operational constraints prevented observations of Arsia Mons by MAVEN after Ls ~260º. While more than 80 Martian sols separate the first and last observations, the AMEC phenomenon is likely to have started earlier (sometime between Ls 235º, when there is a negative observation, and Ls 242º, when there is a positive observation) and to have continued later than Ls 296º, since it was observed in an equivalent season in previous years (Fig. 2).

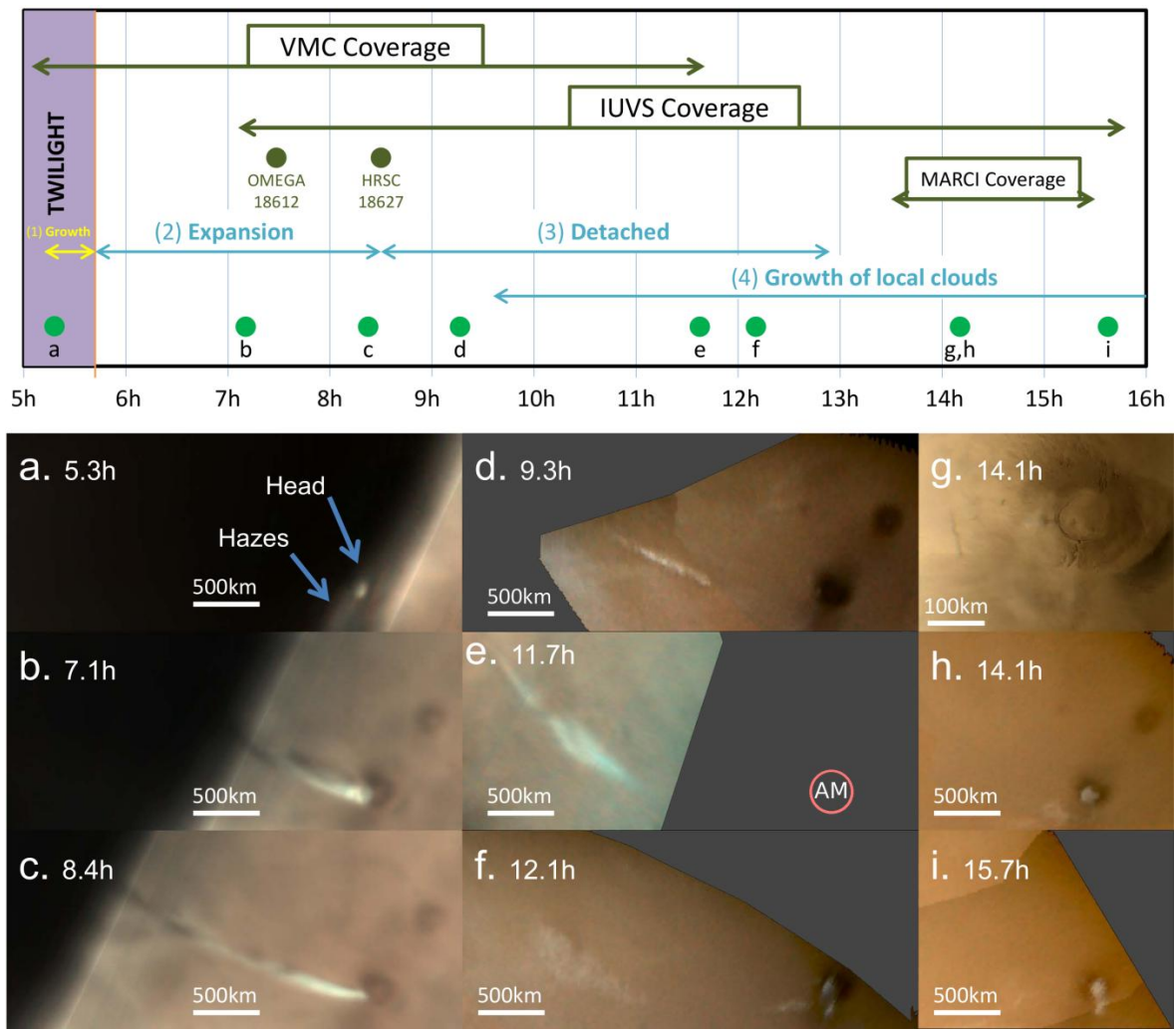

**Figure 5**. Daily cycle of the AMEC in MY34. Top panel: Phases of the cloud evolution and instrument coverage as a function of LTST indicated in Martian hours. Transition time between phases is represented at a fixed time for simplicity, but there is obviously a degree of sol-to-sol variability, see main text for details. Local Times of the images in the bottom panel are indicated by green marks. Bottom panel: examples showing the appearance of the cloud at different local times on cylindrical projections. Gray areas represent non-imaged regions, dark areas correspond to the night. (a) Phase 1. Cloud in twilight, around 30 km above the local areoid. 2018-10-01, Ls 261º. (VMC observation 180340). (b) Phase 2. The cloud is around 1000 km in length. 2018-10-16, Ls 270º. (VMC observation 180362) . (c) Phase 2. The cloud is around 1800 km in length.  2018-10-10, Ls 266º. (VMC observation 180352). (d) Phases 3 and 4. The cloud is detached and thin clouds are visible on the western slope of Arsia Mons:  2018-09-22, Ls 255º. (IUVS, MAVEN orbit 7757). (e) Phase 3. Detached AMEC, Arsia Mons was not imaged and its position is indicated. 2018-10-11, Ls 267º. (VMC observation 180354) . (f) Phases 3 and 4. 2018-09-21, Ls 254º. (IUVS, MAVEN orbit 7752). (g) Phase 4. 2018-10-01, Ls 260º. (MARCI product K09_057091_2603_MA_00N118W). (h) Phase 4. 2018-08-31, Ls 241º. (IUVS, MAVEN orbit 7635). (i) Phase 4. 2018-08-27, Ls 238º. (IUVS, MAVEN orbit 7613).

In what follows, we use the word "tail" to refer to the elongated part of the AMEC, and "head" to refer to the easternmost side of the cloud, next to the volcano where it starts developing. Further details on the morphology of the cloud will be given in section 5.

During the period studied, the AMEC showed remarkable repeatibility in its morphological evolution, following a well-defined daily cycle with different phases that are summarized in Figure 5. We identified the following phases, which we will describe in detail below: (1) Cloud head formation: Before sunrise, the head of the cloud forms and grows in area, probably due to ascending motions (Fig. 5a). As the observation period is around the southern solstice, there were no notable variations of sunrise time at 5.7 LTST, and thus insolation as a function of LTST was practically non-variable during the season. (2) Tail expansion: After sunrise, the cloud rapidly expands westward to generate the distinctive elongated tail (Fig. 5b,c). (3) Detachment and decay: At around 8.3-8.7 LTST, the cloud formation ceases, it detaches from its formation point, and the complete cloud moves westward due to advecting winds until it evaporates before the afternoon. Throughout the rest of the morning, the tail keeps moving to the west whilst slowly fading, probably due to evaporation (Fig. 5d-f). (4) Local clouds: At around 11 LTST, and simultaneously with the decay of the tail in (3), new local clouds, distinct from the AMEC, distributed around the volcano, and smaller in area, begin to form around Arsia Mons. The formation of these clouds always starts on the western slope, and then grows to cover the volcano. A few observations show the presence of local clouds earlier than 11 LTST (for example, in Fig 5d), but they do not grow significantly until 11 LTST. Similar clouds are observed later in the day by MARCI and IUVS (Fig. 5d, 5f-i), and are well known (Benson 2003,2006; Wang & Ingersoll, 2003), but they might not be associated with the occurrence of the AMEC elongated cloud.

### 4.1 Phase 1. Cloud head formation in twilight

VMC images of Arsia Mons taken some minutes before sunrise show the formation of a "twilight cloud" west of the volcano caldera, at ~237º E and 9.5º S (Fig. 5a). The cloud forms during the nighttime, and sunlight starts illuminating the cloud tops before sunrise because of their altitude of 30±5 km above the Martian areoid (Hernández-Bernal et al., 2018). Diffuse hazes extending to the south of the cloud accompany the head during this twilight phase (Fig 5a).

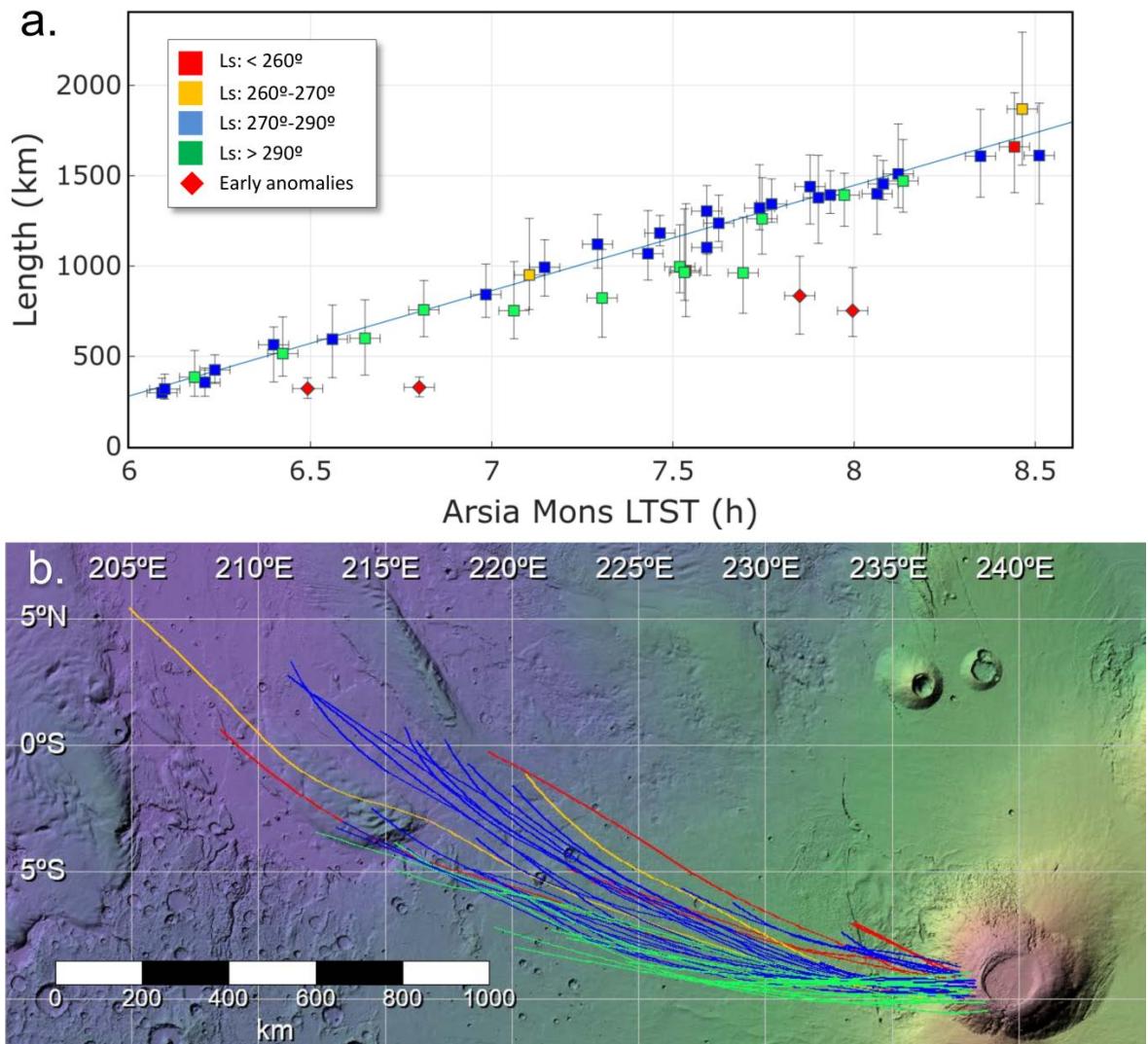

**Figure 6.** Expansion phase of the AMEC. (a) Length of the cloud as a function of LTST. Point incertainties are calculated as explained in Supplementary Text S3 . For long observations, only a representative part of all the available images are represented, separated by ~15min. Linear fit (early anomalies excluded) gives an expansion velocity of 170 ± 10 ms$^{-1}$; (b) Cloud trajectories (traced manually) on a topographic map, the same map as seen in Fig. 4. Color codes indicate the Ls ranges given in the legend in panel (a).

### 4.2 Phase 2. Westward expansion of the tail

For ~3 hours after the sunrise, the cloud grows rapidly in length forming the AMEC phenomenon (Figure 5b,c). We measured the length of the cloud (following its curvature) from available VMC images throughout the season and corresponding to different local times (Figure 6). When a VMC series was long enough, we measured several images in the series, always separated by at least 15 minutes to make sure that variation between images was significant.

Figure 6a shows tail length against local time (from 6 to 8.5 h LTST). Only observations where the whole cloud is visible are included (discarding those images where the cloud penetrates the terminator). The growth is approximately linear, and a linear fit gives an apparent expansion velocity of 170 ± 10ms$^{-1}$ . Assuming a constant rate, and considering the initial size of the head to be around 125km (see section 5.1), the linear fit indicates that the expansion from that size initiates at around 5.8±0.2 LTST (the local sunrise is at 5.7 LTST). The maximum length of the AMEC was measured on October 10 (Ls 266º), with a tail of around 1800 km in length.

In some of the early sols (September 2018), the cloud was shorter than the value expected at the corresponding local time from the general trend, pointing towards some sol-to-sol and/or seasonal variability. These early anomalies are indicated by diamonds in Figure 6a, and could be due to the fact that the GDS 2018 was in its decaying stage throughout that period (Guzewich et al., 2019).

Figure 6b illustrates the extension of the cloud tail at different local times and epochs. The concavity of the curved trajectories is always to the north, suggesting advection of the cloud by persistent combined zonal (westward) and meridional (northward) winds. Due to variations of existing winds, the curvature of the tail varies from sol-to-sol, with local time, and with Ls.

### 4.3 Phase 3. Detachment and decay

At some point around 8.3-8.7 LTST the cloud detaches from its formation point in Arsia Mons. For example, we see the the cloud attached at 8.4 LTST in Fig 5c, and already detached at later local times in Figs 5d and 5e. Some VMC images show that the cloud head narrows before its detachment (Figure 5c, supplementary animation S3), and IUVS observations capture detached clouds after 8.7 LTST (Figure 5d). At the beginning of this phase, there is no remnant cloud activity in the vicinity of Arsia Mons, as the whole cloud moves away from the volcano. Figure 7 shows the distance of the head of the cloud from the location on Arsia Mons where it formed, as measured in IUVS and VMC images. It shows a linear trend, with large dispersion indicating sol-to-sol variability. A linear fit gives a detachment velocity of 150 ± 30 ms$^{-1}$. All observations represented in this graph were acquired before Ls 260º, that is, in the period of the early anomalies mentioned in section 4.2 since due to limited coverage, there are no images of a detached cloud later in the season.

As the sol progresses, both IUVS and VMC observations of the detached cloud reveal a fainter and more turbulent elongated cloud (Figures 5e and 5f). The AMEC eventually disappears, and is no longer visible in IUVS observations at 13 LTST.

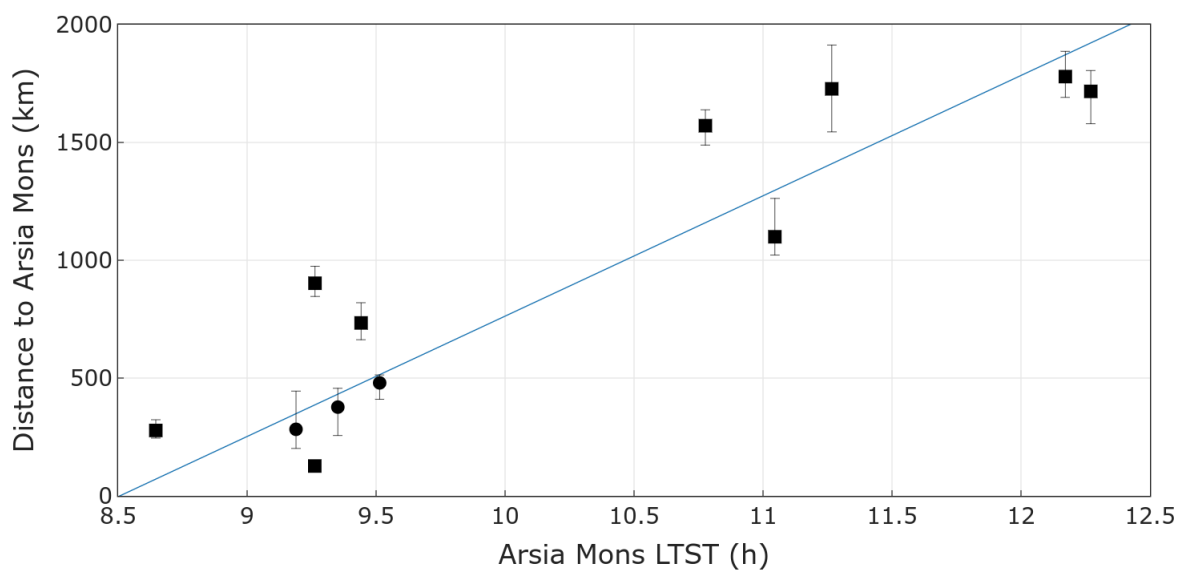

**Figure 7**. Distance of the cloud head to Arsia Mons as a function of LTST during the detachment phase. Point uncertainties are calculated as in Figure 6a. Circles correspond to three measurements within the same VMC observation. Squares represent IUVS observations, each corresponding to a different sol. Due to the coverage limitations, all these observations correspond to the Ls range 240º-260º.

### 4.4 Phase 4. New clouds at Arsia Mons

As the head of the AMEC continues to move away from the base of the volcano, new clouds form over Arsia Mons. A few IUVS images show the development of faint clouds on the western slope of the volcano as early as 9.3 LTST (Figure 5d), but most clouds develop after 11 LTST, and they continue to be present until the last IUVS observation at 15.7 LTST (Figures 5h and 5i). Higher resolution images obtained by MARCI at around 14.0 LTST also show those clouds (Figure 5g). These afternoon clouds appearing around Arsia Mons in this season were also reported in Wang & Ingersoll (2003) and Benson et al. (2003, 2006) and might not be related to the AMEC elongated cloud.

## 5. Morphology and Dynamic Features in MY34

In this section, we describe the morphology of the AMEC, and some further dynamical aspects of the phenomenon.

### 5.1 Position and size of the head of the AMEC before detachment

We measured the position and size of the head of the cloud throughout different local times and sols (Ls) in VMC images (Figure 8a-b). Error was estimated manually, taking into account the limited quality of the images and the diffuse boundaries of the clouds. Approximating the eastern boundary of the head to a semicircle, we determined its diameter and center, and used them as a measure of size and location respectively. Examples of actual head morphologies can be found in Figure 9. The head diameter ranged from 50 to 250 km with a typical value of about 125 km. Because of the dispersion of the data, it is difficult to retrieve a clear temporal evolution of the head size as a function of LTST or Ls (Figure 8c-d). There are hints of a quick head formation early in the morning, reaching a maximum size of ~175 km at ~7 LTST, and then decreasing to around 75-100 km at 8.5 LTST, with no clear trend as the season evolves.

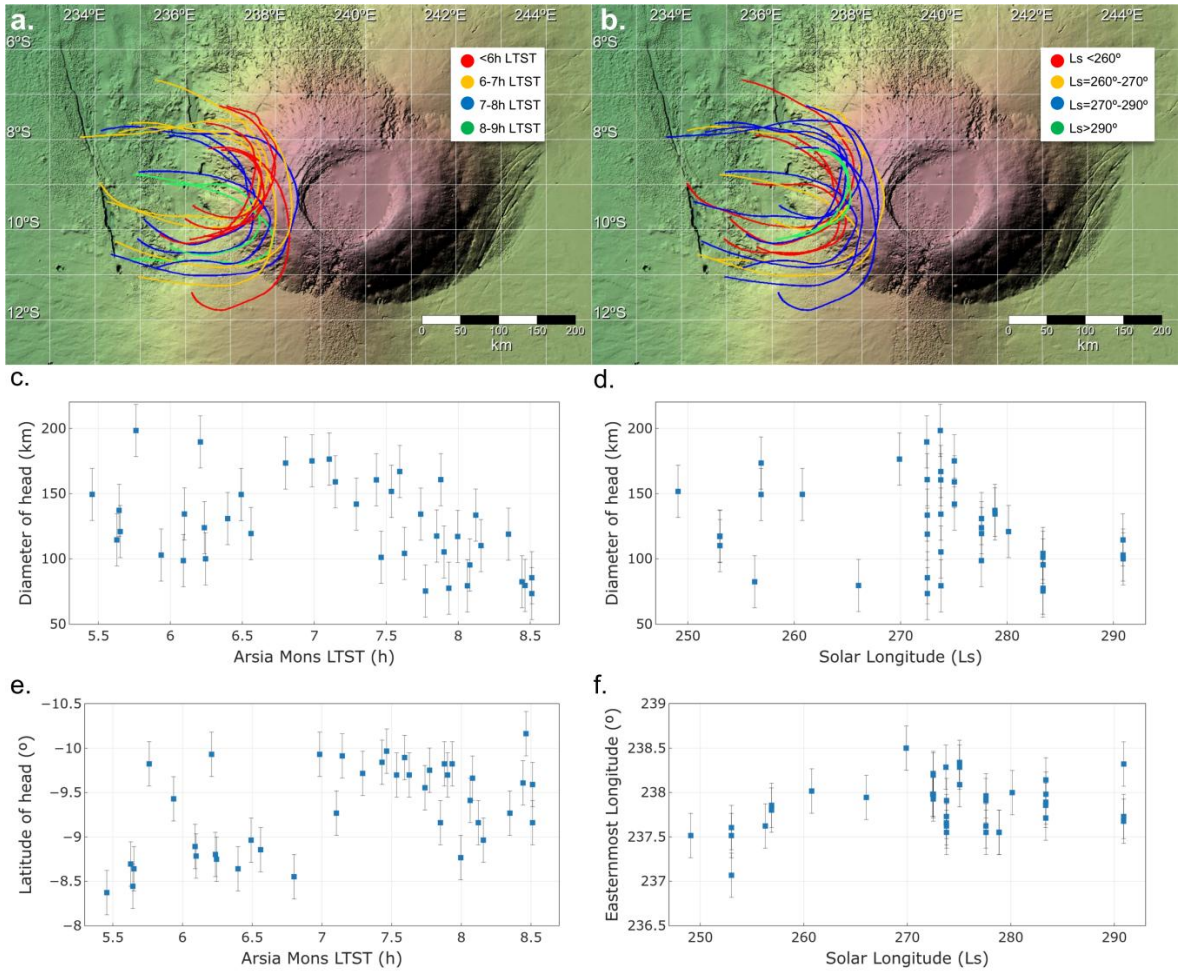

**Figure 8.** Panels a and b: Position and size of the head of the AMEC as a function of LTST and Solar Longitude. The curves (traced manually) represent the head contour and colors identify LTST in (a) and Ls in (b). Panels (c) and (d) Head diameter as a function of LTST and Ls. (e) Latitude of the head center as a function of LTST. (f) Longitude of the head center as a function of Ls.

The central latitude of the head was most often in the range 8.5°-10° S, within the range of latitudes of the Arsia Mons caldera. The easternmost longitude, related to the presence of the cloud on the flank of the volcano, was always lower than 238.5° E, i.e. not reaching the caldera (Figure 8f). The easternmost longitude of the head center occurred in the range Ls ~270°-275°. Although measurements of the position of the cloud might be biased by the effect of parallax, given the estimated height of the cloud (see section 6), this effect is always within the range of error of our measurements.

### 5.2 Morphology of the AMEC during the expansion phase

VMC reveals a rich variety of morphologies, particularly in the head of the cloud, during the expansion phase of the AMEC from 6.5 to 8.5 LTST, when the cloud morphology is already well developed. A scheme of the different features that we observe is presented in Figure 9a.

The earliest occurrences of the AMEC in the season show a small head and a tail with an approximately constant width, similar to the size of the head, and with no signs of turbulence (Figure 9b). This morphology (Type I) was commonly observed before Ls 256º, and different observations differ mainly in the brightness of the tail. Following a gap with no suitable observations for this classification until Ls 266º (see Figure 2), different morphologies are observed at different LTST. In the early hours (~6.9-7.6 LTST) the AMEC shows a Type II morphology (Figure 9c), with a bright textured head and smooth narrowing of the neck into a tail of decreasing width. Later in the day (~7.3-8.4 LTST), we find Type III morphology, with a darker region separating the tail from a still bright head (Figure 9d). This morphology is often associated with a more undulating tail. Finally, prior to detachment (~8.2-8.6 LTST), we often observe Type IV morphology, with a diminishing head and a broad tail that narrows away from Arsia Mons (Figure 9e).

### 5.3 Other features

Although the AMEC can be generally described with the four typologies defined above, there are some variations and extra features worth mentioning.

In some images taken early in the season (Ls 249º and 253º) of appearance of the AMEC phenomenon, the tail of a cloud of Type I morphology is very thin, and the bright head and the dark elongated shadow of the tail dominate the image (Figure 9f). The brightness of the cloud and the visibility of the shadow might be affected by the observation geometry.

Images often show the presence of hazes close to the AMEC. At dawn, the hazes are already visible in the night side of the terminator (see Figure 5a) and they persist into the early hours (see animations in supporting information). Later in the season (after Ls 275º) the AMEC is sometimes accompanied by less diffuse structures, forming cones that extend from its head at an angle of 30º-60º relative to the tail axis (Figure 9g).

In Figure 9h, the tail splits into two sections with slightly different directions (see animation S2 in supporting material), probably due to changes in the intensity and direction of the winds. Finally, the morphology of the tail is sometimes suggestive of bursts of cloud activity in the head advected by winds. Figure 9i shows a ~50 km wide blob detaching from the head to the tail at 8.0 LTST, in a configuration that can be described as type III. This observation continued with a transition to morphology IV, as can be seen in animation S3 in the supporting information.

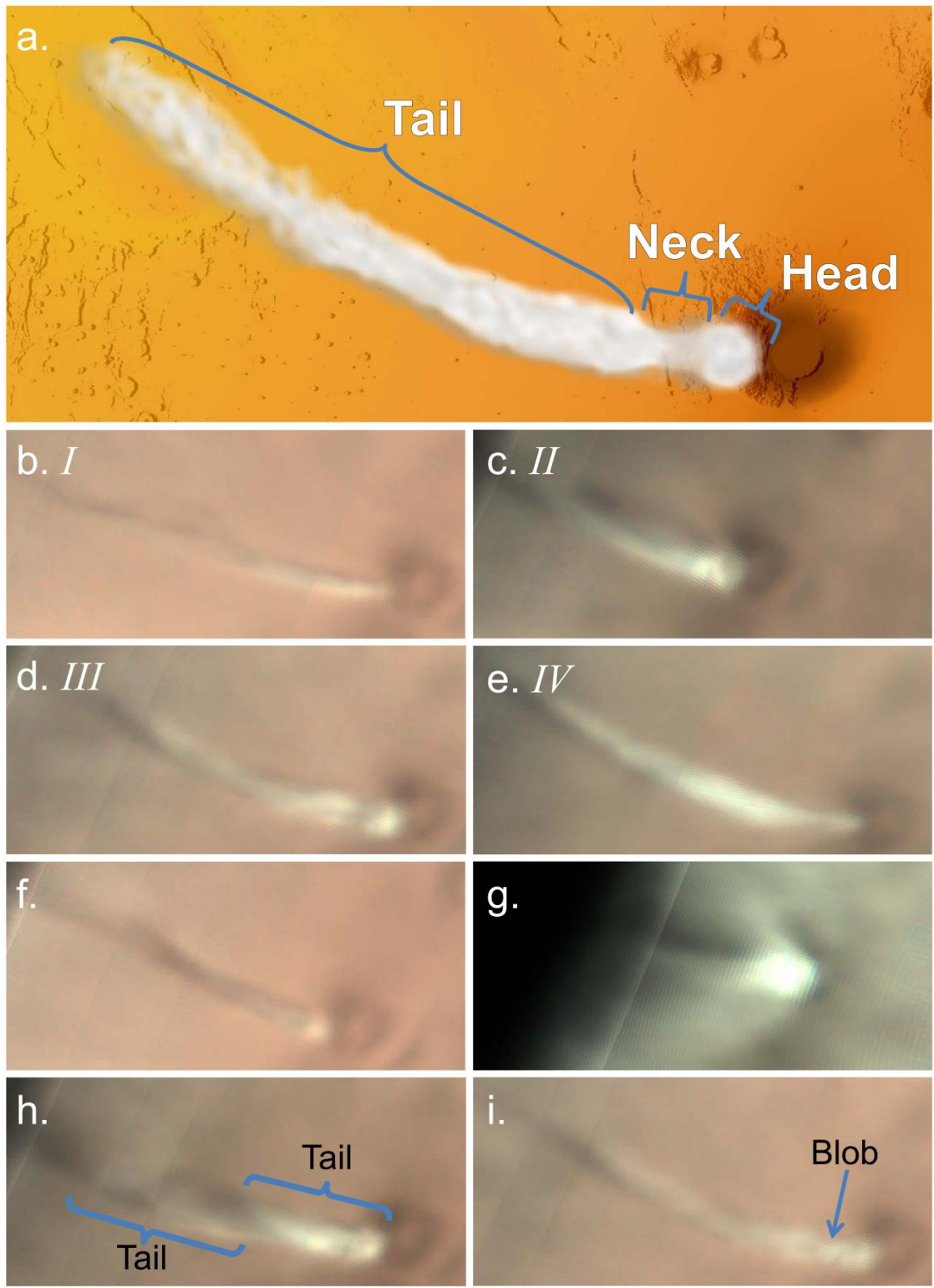

**Figure 9.** Morphology of the AMEC cloud: (a) Scheme of different AMEC features over a synthetic representation, including the head, neck and tail. Panels b-e. Examples of different morphologies. (b) Type I morphology. 2018-09-18, Ls 256º, LTST 8.5. (VMC observation 180326). (c) Type II morphology. 2018-10-18, Ls 270º, LTST 7.1. (VMC observation 180362 ). (d) Type III morphology. 2018-10-20, Ls 272º, LTST 7.9. (VMC observation 180366). (e) Type IV morphology. 2018-11-06, Ls 283º, LTST 8.1. (VMC observation 180397). Panels (f-i) Examples of other features. (f) Bright head with a long dark tail that is probably the shadow projected by the optically thin tail. 2018-09-13, Ls 249º, LTST 7.5. (VMC observation 180306). (g) Hazes in almost symmetrical angles occur ~50º relative to the cloud tail. 2018-11-28, Ls 296º, LTST 6.6. (VMC observation 180424). (h) Two tails at once. 2018-10-22, Ls 274º, LTST 7.4 (VMC observation 180371). (i) Blob in the head being torn off from the head of the cloud (see animation S1 in supporting information). 2018-10-20, Ls 273º, LTST 8.1. (VMC observation 180366).

### 5.4. Evolution throughout MY34

In addition to the marked daily cycle, the AMEC showed a longer-term variability throughout the MY 34. The phases of the daily cycle discussed in section 4.2 were quite regular during the majority of the period studied, with the exception of some early anomalies, marked as red diamonds in Figure 6a. Cloud type I occurs only in this early season, and in these anomalies the cloud usually develops later in the sol, and the cloud length is shorter at any given LTST. For example, on the 19 September there was no cloud at 6.1 LTST, and the cloud was shorter than usual at 8.0 LTST (VMC observations 180317 and 180318). We also find clouds shorter than usual in VMC observation 180329 on the 25 September at 6.9 LTST (Figure 10f), and in the OMEGA observation in orbit 18612 on the 17 September at 7.5 LTST (Figure 3b). This difference in behavior at the beginning of the season was possibly due to the remaining presence of abundant dust from the GDS 2018 (Guzewich et al., 2018).

### 6. Altitude of the cloud and limb observation in MY34

We use three different methods to determine the AMEC altitude:

- Analysis of shadows and illumination conditions.
- Visibility of the cloud tops during twilight (Hernández-Bernal et al., 2018).
- Direct measurement from images of the AMEC phenomenon at the limb of Mars.

The observations used in the analysis of the cloud altitude in MY34 are presented in Figure 10. All altitudes in this section are relative to the Martian areoid as defined by MOLA data. The accuracy of these measurements is highly dependant on how sharp the border of the clouds is, as detailed in supporting material text S4.

Figures 10a and 10f show the shadow projected by the cloud in two different stages of the phenomenon and with different illumination conditions. Figure 10a shows the cloud and its shadow as seen in high resolution by HRSC. We studied this shadow in three dimensions, taking into account the parallax effect due to the high altitude of the cloud, the position of the observer, and the local topography. We found that the tail of the cloud was around 40-50 km in altitude. The profile of heights along the tail is indicated by the blue line in Figure 10g. The uncertainty of these measurements is mostly due to the difficulty of determining the limits of the cloud and the shadow, because of the presence of surrounding hazes. It is interesting to note that, within measurement uncertainties, the head of the cloud is higher than the tail, and the farthest extreme of the tail is at a lower altitude.

Figure 10f is a low-resolution image acquired by VMC taken on a sol when the expansion of the cloud was delayed with respect to the trend established in section 4 (an early anomaly). Local time at Arsia Mons is 6.8 LTST and the low solar elevation angle (~15°) results in a long shadow that extends to the terminator. Simple geometry calculations indicate a minimum altitude of 45 km for the cloud tops and a vertical extension of the cloud of at least 5-10 km. The extension and altitude of this cloud is represented by the green area in Figure 10g.

Figures 10b and 10c shows images obtained with VMC of the AMEC cloud at the limb of Mars.  The curves corresponding to the limb in these images are shown with white lines in Figure 10d. The limb observation shows aerosols around Arsia Mons and in its eastern side. Note that this observation was performed at Ls 281°, a season where hazes are also commonly observed in nadir observations of the region (section 5.3). On the eastern side of Arsia Mons (yellow arrows in Fig. 10c), optically thick aerosols are present up to an altitude of 35±10 km, and an optically thinner detached layer is present at altitudes of 45±10 km. On the western side of the volcano (cyan arrows in Fig. 10c), the head appears at a slightly higher altitude next to Arsia Mons (~50 ±10 km) in agreement with measurements of the shadow in the HRSC image (see Fig 10g for comparison).

Figure 10e is a good example of a twilight observation by VMC. From an analysis of image sequences like this one, we constrained the top altitude of the cloud to be 30±5 km.

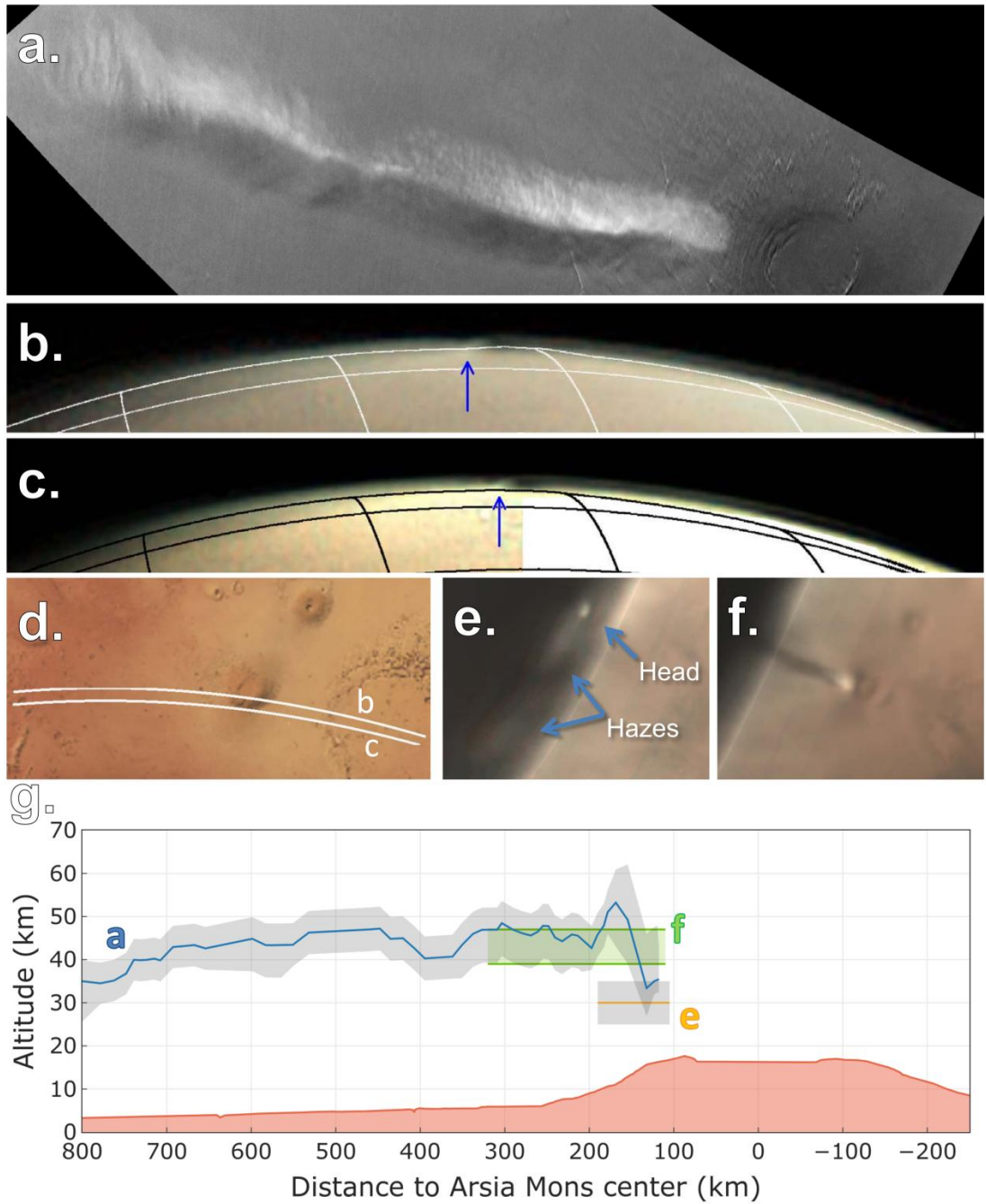

Finally, figure 10g summarizes the results of our altitude measurements. Our results suggest that the cloud head raises during twilight, when we measure lower heights for the head cloud tops than at the start of expansion.

**Figure 10**. Altitude of the AMEC relative to the areoid from different observations. (a) Shadow of the cloud at high resolution as observed by HRSC. This image has been enhanced to improve shadow visibility, for the equivalent color image refer to Fig. 3d. Acquired on 2018-09-21, Ls 254º, LTST 8.5, (orbit 18627). (b,c) AMEC cloud seen at the limb, clearly visible on the western flank (left of the volcano position, indicatedwith dark blue arrows), with hazes and a

**An Extremely Elongated Cloud over Arsia Mons Volcano on Mars: I. Life Cycle.**
Hernández-Bernal et al. 2020. Manuscript accepted for publication on Journal of Geophysical Research
This document is distributed under CC BY-SA 3.0 IGO licensedetached layer of aerosols east of Arsia Mons (right of the blue arrow). These images were acquired 48 seconds apart on 2018-11-03, Ls 281°, LTST 6.5, (VMC observation 180390). (d) Map of the region showing limb trajectories for b and c panels. (e) Head of the cloud imaged during twilight. Hazes also appear around the head in twilight (indicatedwith arrows). 2018-10-01, Ls 261°, LTST 5.3, (VMC observation 180340). (f) Shadow of the cloud shortly after dawn. 2018-09-25, Ls 257°, LTST 6.9, (VMC image 180329). (g) Graph showing the topographic profile (red profile at the bottom) as well as the altitude of the cloud shown in panels (a) (blue line) and (e) (orange line), where gray areas indicate the uncertainty in altitude. Also shown is the altitude and minimum thickness of the cloud shown in panel (f).

## 7. Interannual comparison

We searched for the AMEC phenomenon in a survey of available images taken by Mars orbiters. We only found suitable observations (in the expected ranges of LTST and Ls) in MYs 29 to 33 (see Figure 2). We also include an observation in MY12, corresponding to the Viking 2 spacecraft. Poor coverage of earlier observations does not allow us to reveal the full seasonal cycle of the cloud appearance, but a conservative estimation from our observations is that it takes place between Ls = 220°-320°, although its lifespan might be longer (see Figure 2).

Figure 11 shows examples of images of the cloud in different MYs and in the following subsections we explore potential interannual variability of the AMEC features.

When comparing the onset date of the phenomenon for the available MYs (see Figure 2), we see that the earlier appearances were in MY31 (Ls 226°) and MY32 (Ls 233°). In MY34 we have suitable VMC observations at these Ls that do not show the AMEC phenomenon (see Figure 2 and Table S1), suggesting that the onset of the AMEC season that year was delayed in comparison with MYs 31 and 32. This delay is possiblyrelated to the occurrence of the MY34 GDS, and the progressive clearingof dust in the atmosphere (Guzewich et al., 2019; Sánchez-Lavega et al., 2019).

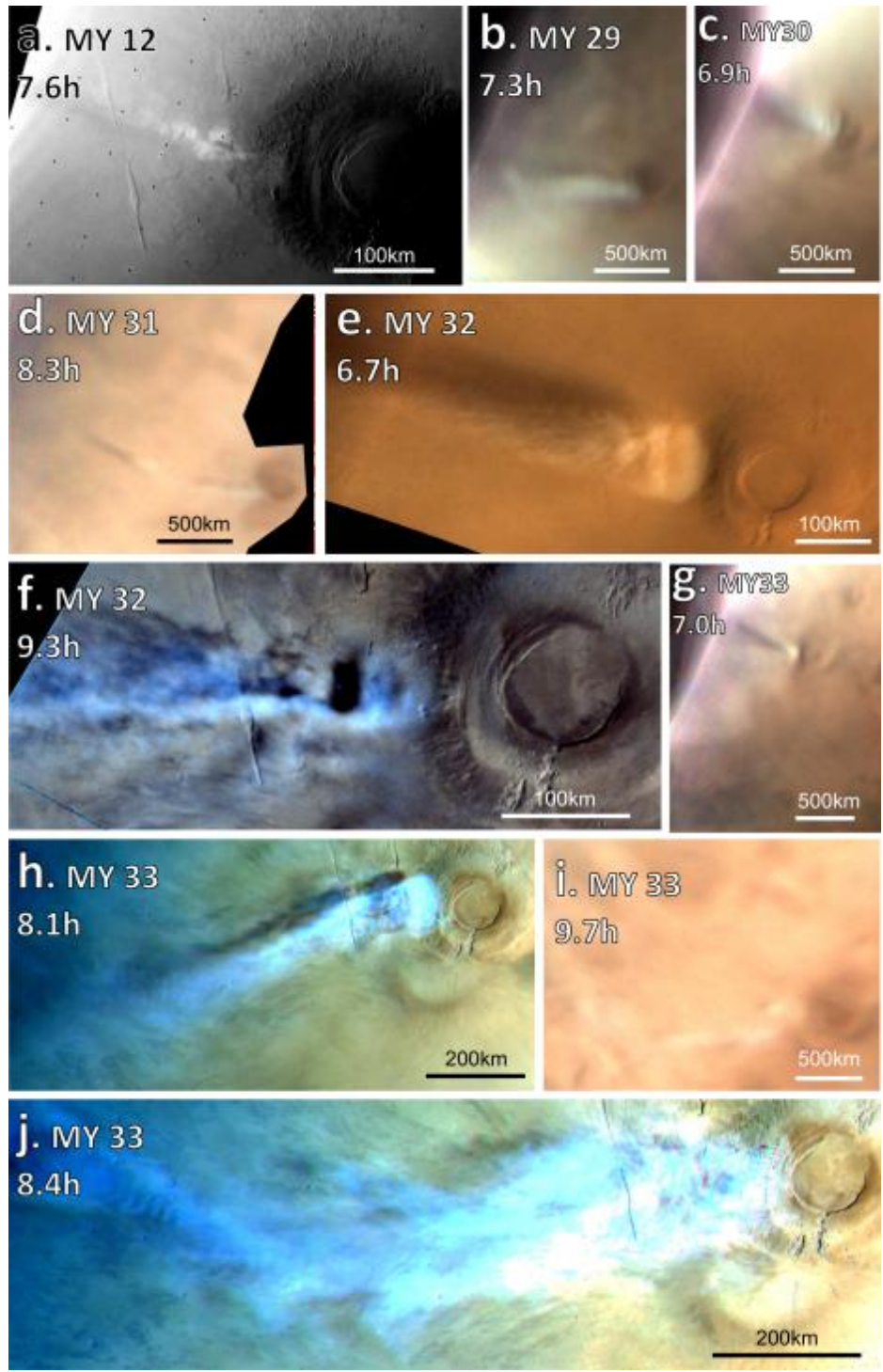

**Figure 11.** Images of the AMEC phenomenon seen in different MYs. All images are projected onto cylindrical maps. Spatial scale is indicated in each panel. (a) MY 12. 1977-08-11, Ls 314° (Viking Orbiter 2 observation 356b33). (b) MY 29. 2009-07-2, Ls 296° (VMC observation 090025). (c) MY 30. 2011-03-21, Ls 258° (VMC observation 110005). (d) MY31. 2012-12-15, Ls 226° (VMC observation 120060). (e) MY 32. 2014-12-13, Ls 252° (MCC product MCC_MRD_20141213T134802312_D_GDS). (f) MY32. 2015-01-04, Ls 266° (MCC product

MCC_MRD_20150104T063053645_D_D32). (g) MY 33. 2016-12-25, Ls 287º, (VMC observation 160150). (h) MY33. 2017-01-05, Ls 294º, (HRSC observation in orbit 16486) (i) MY33. 2017-01-08, Ls 296º (VMC observation 170010). (j) MY 33. 2017-01-01, Ls 291º (HRSC observation in orbit 16472).

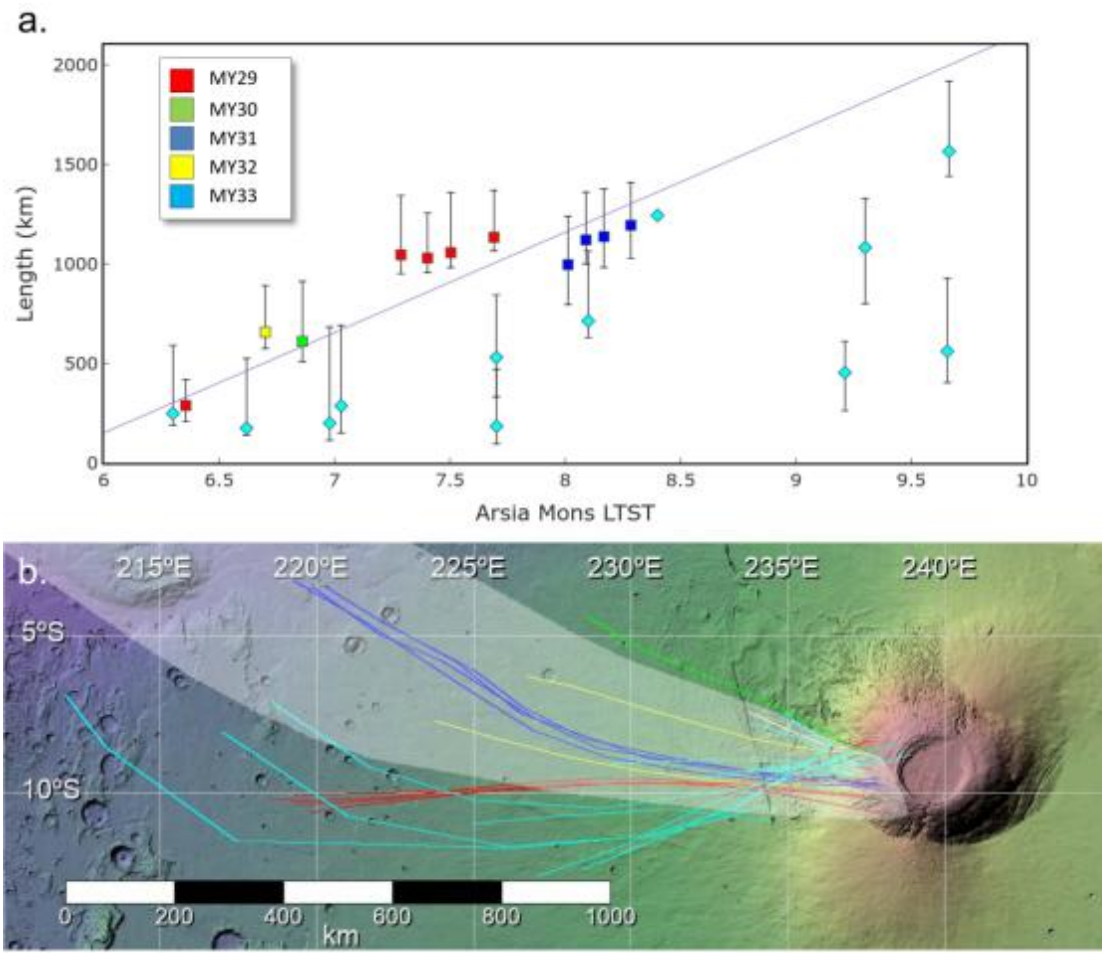

**Figure 12.** Length and Curvature of the AMEC in MYs 29-33 (Analogous to Figure 6). Panel a. Length of the AMEC as a function of local time. Color code indicates MY. The linear fit corresponds to all data except for MY 33 (indicated with diamond shaped markers),. Panel b. Outline of the tail of the cloud over a topographic map as in Figure 6b. The area occupied by the cloud trajectories in MY 34 (Fig. 6b) is indicated by a background white shadow for comparison.

### 7.1 Daily Cycle: Expansion and detachment

Figure 12a shows the length of the AMEC as a function of LTST for MYs 29-33, measured only in the images where the complete cloud was visible. The global trend of growth with LTST is similar in MYs 29-32, and a linear fit to data corresponding to those four years indicates an expansion velocity of 130 ± 20 ms$^{-1}$, slower than in MY34. The fit is not as good as in Fig. 6 (which contained data from MY 34 only), as it must reflect a degree of interannual variability. The trend in MY 33 is different, and the AMEC is often considerably shorter in length than expected from the general trend in other years, resembling the "early anomalies" reported for MY 34 in section 4.2 (e.g. compare Fig. 11g with Fig. 10f). The expansion direction of the cloud is similar to MY34, initially to the west, and curving towards the north (Figure 12b). An exception to this is MY 33, when the cloud expanded initially southward of Arsia Mons, but keeping its northward pointing concavity.

Few observations in previous years were acquired after 9 LTST and before noon, the local time around which the AMEC had always detached from Arsia Mons in MY 34. MY 33 is an exception to this, as VMC observed past 9 LTST on several occasions. Unlike in MY34, in images before 10 LTST, the cloud had not clearly detached from Arsia Mons, and no elongated cloud was found between 10 and 12 LTST.

### 7.2 AMEC morphology

In general, morphologies of the cloud in all years are similar to those reported in section 5.2 for MY34. For example, Figure 11c shows a morphology resembling types II or III discussed in section 5.2; Figure 11d shows a clear case of type I and Figure 11e resembles type II. Figure 11h corresponds to type III, indicating that the dark feature usually observed in the neck in low-resolution images of this morphology is the result of a lack of thick clouds in the area. Figure 11g closely resembles the early anomaly depicted in Figure 10f. Figures 11f and 11i have no clear equivalent in MY 34, and it is worth mentioning that both were obtained at a local time when the cloud was usually detached in MY 34.

Hazes and conical shapes such as those reported in section 5.3 are also found in VMC images from the previous years, as in Fig. 11b and 11c, which show hazes southward of the AMEC. The high-resolution images in Fig. 11f,h show hazes southward of the elongated cloud, and extended hazes in the west in the vicinity of the terminator.

If we concentrate on specific features of the AMEC observed at high resolution, the sol-to-sol variability prevails over the seasonal evolution, as is apparent in Figs. 11h,j, obtained by HRSC only four sols apart and almost at the same local time.

### 7.3 Cloud altitude in previous years

We do not have any limb or twilight observations from previous years, and thus all our estimations of height are obtained by analyzing the shadow of the cloud in high-resolution images. The HRSC image obtained in orbit 16486 in MY 33 (Fig. 11h) shows the AMEC casting a prominent shadow at 8.1 LTST that gives a cloud altitude of 40-45 km above the areoid. The MCC observations of the shadow at Ls 252º(Fig. 11e) at 6.7 LTST gives an altitude of 34±3 km above the areoid. Another MCC observation on 4 January 2015 (Fig. 10f) at 9.2 LTST shows the projected shadow from the head from which we derive an altitude of 45±2 km. All these altitudes are in very good agreement with those reported previously in section 6 for MY 34.

## 8. Discussion and Conclusions

In this paper, we report the formation of an extremely Elongated Cloud in the early hours of the morning, extending to the west from the Arsia Mons volcano (AMEC, Arsia Mons Elongated Cloud). We analyze several features of this cloud phenomenon and its changes as a function of Local Time, Solar Longitude, and MY. In our study, we have focused on the occurrence of the AMEC phenomenon during MY 34, which started during the decaying phase of a Global Dust Storm (GDS 2018), and we have also presented observations of the phenomenon in previous MYs.

Morphologically, the AMEC has two main regions (Fig. 9a): the head, a bright cloud approximately round in shape but with texture, of an average size of ~125 km in diameter, that forms above the western flank of the volcano at longitude 237ºE and latitude 8.5ºS-10ºS; and second the long tail that grows and expands westward toward the terminator during the morning hours at speeds of around 170 ms$^{-1}$ in MY 34 (slower in other years) The elongated cloud attains a maximum length of 1800 km but its width typically remains below 200 km. The head reaches maximum altitudes of 40 – 50 km above the areoid. We find that the AMEC follows a daily cycle with head formation around sunrise(5.7 LTST), followed by its expansion and detachment from the volcano point origin at ~8.5 LTST. In the following hours, the cloud becomes thinner and more turbulent, evaporating whilst new thin clouds grow locally over Arsia Mons, as observed in the afternoon in previous MYs (Wang & Ingersoll, 2002; Benson et al., 2003; 2006).

Due to the fact that the AMEC phenomenon occurs at the Martian sunrise, we have few observations that show the details of the head formation in high resolution. One of the best images of the earlier head morphology was obtained by Viking Orbiter 2 in MY12 (Ls 308°) (Fig. 5 in Hunt and Pickersgill, 1984). The image, obtained at LTST 6.6, shows a high dynamical richness in the head, and another image (Fig. 6 in the same reference) at Ls 314º shows the plume growing some hundred kilometers to the west of Arsia Mons at 7.6 LTST, in good agreement with what we have observed in recent years.

The pattern traced by the AMEC during its development and expansion typically exhibited a curved shape with its concavity pointing toward the north (Figs. 6 and 12), as driven by the ambient zonal and meridional winds. We found evidence suggesting seasonal changes and interannual variations, but we also observed strong sol-to-sol variations. Thus, the tail length and orientation are a good proxy to characterize the ambient wind field structure and its variability.

With these data in hand, the scenario we propose is that the cloud head formation takes place when moist air is forced to ascend in updrafts along the volcano slopes, and the head forms when water ice condenses due to low temperatures at higher altitudes. We suggest that after the head formation, high altitude westward winds expand and drag the cloud to the west. The vertical ascent of air parcels into the Martian mesosphere driven by anabatic winds along the slopes of the giant Mars volcanoes has been studied by means of mesoscale models (Rafkin et al., 2002; 2017; Michaels et al., 2006). These studies show that water vapour may be transported up to altitudes higher than 40km (Michaels et al., 2006). On the other hand, recent works have shown that the hygropause can reach up to 90 km during a GDS (Heavens et al, 2018). In a second paper, we will present a full theoretical study of the AMEC phenomenon based on its orographic nature and numerical models.

Observations in the afternoon show that Arsia Mons is the only location on Mars where water ice clouds are usually found in the season of occurrence of the AMEC (Benson et al. 2003, 2006; Wang & Ingersoll 2002). The AMEC takes place in the early morning and thus it was not part of the mentioned studies. Previous works have studied the water vapor concentration on the slopes of the Tharsis giant volcanoes (Maltagliati et al. 2008; Titov et al., 1994). This topic together with other aspects regarding the AMEC formation will be discussed in the second part of this work.

The AMEC phenomenon develops during the season of maximum insolation at the volcano (daily average at the Top of the Atmosphere TOA ranges from ~ 200 to 225 Wm$^{-2}$; see Fig. 3 in Sánchez-Lavega et al. 2019) with the perihelion occurring at Ls 251° and the southern solstice at Ls 270°. The AMEC season is also the period when dust injection in the atmosphere and the development of dust storms on Mars take place (Kahre et al., 2017; see also bottom panel of Fig. 3, in Sánchez-Lavega et al., 2019). Therefore, the two phenomena, the AMEC and dust (via GDS and regional storms) occur simultaneously, making the AMEC a good case study for the effect of dust on the dynamics and distribution of water vapor, and the microphysics of orographic cloud formation.

There is possible evidence of the effects of dust storms on afternoon cloud development at Arsia Mons. Benson et al. (2003) reported no afternoon cloud activity in Arsia Mons between Ls 227°-235° during MY 24 at the time of a regional dust storm (Cantor et al., 2001). Later, Benson et al. (2006) also reported less cloud formation over Arsia Mons during the GDS 2001 that took place in MY 25 (Cantor, 2007; Montabone 2015; Strausberg et al., 2005). We have explored the AMEC cases in MYs with Global Dust Storm (MY12, MY34) and without (MYs 29-33). We use the earliest AMEC observations among all the MYs studied, which corresponds to Ls 226º in MY 31 (December 2012), as a reference. The Viking Orbiter images of the AMEC described above (MY12, Ls 307° and 317º, July 1977) were obtained during the GDS 1977B that took place between May and October 1977 (Ls 268° - 349°; Pollack et al., 1979), indicating that both phenomena coexisted. In MY 34, we find a faster expansion compared to previous years, and a seasonal delay in the Ls of the onset of the AMEC phenomenon, compared to MYs 31 and 32, and some "early anomalies" occurring at the beginning of the season (captured at Ls 240º-260º), consisting of a later formation of the cloud head and a shorter development of its tail. These features are possibly related to the fact that the onset of the AMEC coincided with the decline of the GDS 2018 (Guzewich et al., 2019; Montabone et al., 2020; Smith 2019b).

How rare is the AMEC phenomenon? The presence of the Valles Marineris Cloud Trails (VMCT) reported by Clancy et al. (2009; 2014), strongly resembles the AMEC. These VMCT are observed recurrently each MY in the Valles Marineris and Noctis Labyrinthus areas around the perihelion season (Ls around 230º-260º) in the afternoon hours (13-15 LTST). Observed in MY28, the clouds were quite elongated, with lengths of 400-1000 km and widths of 25-75 km, and altitudes as deduced from shadows of around 40-50 km, with the cloud expansion requiring winds over 100 ms$^{-1}$ (Clancy, 2009). Another feature that might be similar to the AMEC is an elongated cloud that has been reported over Ascraeus Mons (the northern Tharsis volcano), first observed by the Viking orbiters (Carr et al., 1980, figure 141).

Future observations of the AMEC should focus on detailed and systematic images by spacecraft in non-sun-synchronous orbits in order to track the evolution of the phenomenon as a function of local time and, in the long-term, to monitor its interannual variability (current available orbiters are MEX, MAVEN, MOM and TGO). Wide angle cameras like VMC and MCC are well suited to cover large areas at once and obtain sequences of images that enable the study of the phenomenon globally as a function of LTST, and the EXI instrument onboard the Emirates Mars Mission (EMM) will also have this capability. Some pushbroom scanners, like IUVS or HRSC, can also obtain regular images of large spatial areas that could show this kind of phenomenon. Narrow angle cameras can give details of the cloud morphology at its onset, and spectral imagers (e.g. MEX/OMEGA, TGO/CASSIS, EMM/EXI) and other instruments (e.g. MEX/PFS) can give information on cloud particle sizes and optical depths and retrieve temperatures and water vapor abundances. A good characterization of dust in the atmosphere can help in understanding the connection between dust storms and the varied behavior of the AMEC. All these parameters will enable the future modelling of this outstanding Martian meteorological phenomenon.

## Acknowledgments, Samples, and Data

This work has been supported by the Spanish project AYA2015-65041-P and PID2019-109467GB-I00 (MINECO/FEDER, UE) and Grupos Gobierno Vasco IT-1366-19. JHB was supported by ESA Contract No. 4000118461/16/ES/JD, Scientific Support for Mars Express Visual Monitoring Camera. The Aula EspaZio Gela is supported by a grant from the Diputación Foral de Bizkaia (BFA). We acknowledge support from the Faculty of the European Space Astronomy Centre (ESAC). Special thanks are due to the Mars Express Science Ground Segment and Flight Control Team at ESAC and ESOC. The contributions by K.C and N.M.S were supported by NASA through the MAVEN project.

A list of observations used in this paper is provided in the Supporting Information. All observations are available in ESA PSA, NASA PDS, and ISRO ISSDC. Specific links are given in the supporting material. We uploaded the images as they were used in this study to a repository publicly available: https://doi.org/10.6084/m9.figshare.c.5212793

We acknowledge Claus Gebhardt and an anonymous reviewer whose comments helped improve and clarify this manuscript.

We acknowledge the use of data from the Mars Orbiter Mission (MOM), first interplanetary mission of the Indian Space Research Organization (ISRO), archived at the Indian Space Science Data Centre (ISSDC).

**An Extremely Elongated Cloud over Arsia Mons Volcano on Mars: I. Life Cycle.**
Hernández-Bernal et al. 2020. Manuscript accepted for publication on Journal of Geophysical Research
This document is distributed under CC BY-SA 3.0 IGO licenseSánchez-Lavega, A., Chen-Chen, H., Ordonez-Etxeberria, I., Hueso, R., del Rio-Gaztelurrutia, T., Garro, A., & Wood, S. (2018a). Limb clouds and dust on Mars from images obtained by the Visual Monitoring Camera (VMC) onboard Mars Express. Icarus, 299, 194-205.

Sánchez-Lavega, A., Garro, A., Río-Gaztelurrutia, T., Hueso, R., Ordóñez-Etxeberria, I., Chen Chen, H., et al. (2018b). A seasonally recurrent annular cyclone in Mars northern latitudes and observations of a companion vortex. Journal of Geophysical Research: Planets, 123, 3020–3034.

Sánchez-Lavega, A. Del Río-Gaztelurrutia, T., Hernández-Bernal, J., & Delcroix, M. (2019). The Onset and Growth of the 2018 Martian Global Dust Storm. Geophysical Research Letters, 46, 6101-6108. https://doi.org/10.1029/2019GL083207

Slipher, E. C. (1927). Atmospheric and surface phenomena on Mars. Publications of the Astronomical Society of the Pacific, 39(230), 209-216.

Smith, S. A., & Smith, B. A. (1972). Diurnal and seasonal behavior of discrete white clouds on Mars. Icarus, 16(3), 509-521.

Smith, D. E., Zuber, M. T., Frey, H. V., Garvin, J. B., Head, J. W., Muhleman, D. O., ... & Banerdt, W. B. (2001). Mars Orbiter Laser Altimeter: Experiment summary after the first year of global mapping of Mars. Journal of Geophysical Research: Planets, 106(E10), 23689-23722.

Smith, M. D. (2019a). Local time variation of water ice clouds on Mars as observed by THEMIS. Icarus, 333, 273-282.

Smith, M. D. (2019b). THEMIS observations of the 2018 Mars global dust storm. Journal of Geophysical Research: Planets, 124(11), 2929-2944.

Strausberg, M. J., Wang, H., Richardson, M. I., Ewald, S. P., & Toigo, A. D. (2005). Observations of the initiation and evolution of the 2001 Mars global dust storm. Journal of Geophysical Research: Planets, 110(E2).

Sutton, J. L., Leovy, C. B., & Tillman, J. E. (1978). Diurnal variations of the Martian surface layer meteorological parameters during the first 45 sols at two Viking lander sites. Journal of the atmospheric sciences, 35(12), 2346-2355.

Wang, H., & Ingersoll, A. P. (2002). Martian clouds observed by Mars global surveyor Mars orbiter camera. Journal of Geophysical Research: Planets, 107(E10), 8-1.

Wellman, J. B., Landauer, F. P., Norris, D. D., & Thorpe, T. E. (1976). The Viking orbiter visual imaging subsystem. Journal of Spacecraft and Rockets, 13(11), 660-666.

Wilson, R. J., Neumann, G. A., & Smith, M. D. (2007). Diurnal variation and radiative influence of Martian water ice clouds. Geophysical Research Letters, 34(2).36